\documentclass[prd,a4paper,aps,epsfig,showpacs,twocolumn,floats]{revtex4}

\usepackage{amssymb,amsmath,hyperref}

\newcommand{\be}{\begin{equation}}
\newcommand{\ee}{\end{equation}}
\newcommand{\bea}{\begin{eqnarray}}
\newcommand{\eea}{\end{eqnarray}}

\newcommand{\db}[1]{\overline{\delta{#1}}}
\newcommand{\Ha}{\mathcal{H}}
\newcommand{\Ga}{\mathcal{G}}
\newcommand{\ep}{\varepsilon}
\newcommand{\pr}{^{'}}
\newcommand{\dbu}{\db{u}_{\parallel}}
\newcommand{\ct}{{c_s}^2}
\newcommand{\cs}{{c_s}^2}
\newcommand{\xx}{\xi_{ad}}
\newcommand{\xz}{\zeta_{ad}}
\newcommand{\xit}{\tilde{\xi}}
\newcommand{\zetat}{\tilde{\zeta}}
\newcommand{\ba}{\begin{align}}
\newcommand{\ea}{ \end{align} }
\newcommand{\bs}{\begin{split}}
\newcommand{\es}{\end{split}}

\newcommand{\pp}{^{''}}

\newcommand{\Psp}{|{\mathcal P}_{\theta} (k)|^2}

\newcommand{\zfpp}{\frac{z\pp}{z}}

\newcommand{\zfpt}{z\pr / z}
\newcommand{\afpt}{a\pr / a}
\newcommand{\ffpt}{f\pr / f}
\newcommand{\zfppt}{z\pp / z}

\newcommand{\gc}{\frac{2}{3 \gamma}}
\newcommand{\pa}{\partial}
\newcommand{\ce}{c_{es}^2}

\begin{document}

\title{Non-adiabatic primordial fluctuations}
\author{Johannes Noller and Jo\~ao Magueijo}
\affiliation{
Theoretical Physics Group, Imperial College, London, SW7 2BZ}
\date{\today}
\begin{abstract}
We consider general mixtures of isocurvature and adiabatic cosmological perturbations.
With a minimal assumption set consisting of the linearized Einstein equations and a primordial perfect fluid
we derive the second-order action and its curvature variables. We also allow for varying equation of state and speed of sound profiles. The derivation is therefore carried out at the same level of generality that has been achieved for adiabatic modes before. As a result we find a new conserved super-horizon quantity and relate it to the adiabatically conserved curvature perturbation. Finally we demonstrate how the formalism can be applied by considering a Chaplygin gas-like primordial matter model, finding two scale-invariant solutions for structure formation.
\end{abstract}
\pacs{0000000}
\maketitle

\bigskip

\section{Introduction}

Cosmology, as every other field in science, comes equipped with its set of paradigms. Here we wish to break with the assumption that cosmological perturbations are adiabatic. Departure from this claim has of course already been discussed and considered in many forms in the literature (see e.g. \cite{Trotta:2004sm, Bean:2006qz, Hikage:2009rt}).
The {\it novelty} of this paper then is to develop the cosmological perturbation formalism in a relativistic context for any mixture of entropy and adiabatic perturbations in the same generality that has been achieved for purely adiabatic perturbations before (see e.g. \cite{mukh}). As such we will carry out our derivations employing a {\it minimal assumption set} only consisting of 
\begin{itemize}
\item the linearized Einstein equations and 
\item modelling primordial matter as a perfect fluid. 
\end{itemize}
Importantly we therefore do not constrain the equation of state or speed of sound profile of the primordial fluid, which are consequently allowed to vary freely.

In a nutshell our philosophy is therefore to say as much as possible about non-adiabatic perturbations whilst specifying as little as possible w.r.t. the nature of the underlying primordial matter. The resulting formalism will allow straightforward computation and comparison of e.g. power spectra produced by any model satisfying our assumption set. In contrast, previous literature has focussed on considering isocurvature perturbations produced by more tightly constrained matter in great detail (e.g. matter-radiation and other fluid mixtures~\cite{kodama,mukh,lidsey} or multiple scalar fields~\cite{Langlois:2008mn}).

Observationally, purely adiabatic modes are still compatible with current experimental data, but near-future experiments such as Planck should soon tighten constraints, possibly establishing the presence of entropic modes. In fact there might already be positive evidence suggesting their presence (see e.g. \cite{Valiviita:2009bp}).

The paper is organized as follows. In section \ref{Nad} we give a precise definition of what we mean by ``non-adiabatic'' perturbations and give explicit examples to show how isocurvature modes can arise from the interaction of components in an overall perfect fluid.
In \ref{lin} we then extract equations of motion from the linearized Einstein equations for a perfect fluid. 
These are developed into a second-order action for fluctuations and the associated curvature variables in the adiabatic case in \ref{adia}. Here we also show that single scalar fields generically have vanishing non-adiabatic perturbations on large scales.
In section \ref{nadia} we then present the main results w.r.t. the formalism developed in this paper: A second-order action for fluctuations and associated curvature variables in the non-adiabatic case. We pay special attention to the asymptotic limits of the solution, finding a new conserved super-horizon charge.

Having mentioned some observational constraints imposed on primordial matter models in \ref{horiz}, in section \ref{model} we provide an application of the formalism developed. More specifically we investigate a Chaplygin gas-like primordial matter model, finding two solutions capable of producing scale-invariant power spectra and resolving the horizon problem.
Finally we look at the issue of growing and decaying modes in \ref{pertrev}, before concluding in section \ref{conc}

\section{(Non-)Adiabaticity} \label{Nad}

\subsection{The hydrodynamical setup}

Let us begin by spelling out what exactly we mean by ``non-adiabatic perturbations''. In staying with our philosophy of leaving the exact nature of the primordial matter in question unspecified, the formalism developed here will be at the level of hydrodynamics and we will thus characterize the system thermodynamically by its pressure ($p$), energy density ($\ep$) and entropy ($S$) variables. When Taylor-expanding perturbations in the pressure in terms of the other variables we find
\bea
\delta p &=& \left.\frac{\pa p}{\pa \ep} \right|_S \delta \ep + \left.\frac{\pa p}{\pa S}\right|_{\ep} \delta S \\
&=& \ct \delta \ep \hspace{.5cm} + \hspace{.5cm} \delta p_{ent},
\eea
where we have denoted the entropic perturbation term as $\delta p_{ent}$ and have defined the adiabatic speed of sound $\ct$ as
\be
\ct = \left.\frac{\pa p}{\pa \ep} \right|_S = \frac{\dot{p}}{\dot{\ep}}.
\ee
where $\dot{}$ is a derivative w.r.t. proper time $t$.

Our demarcation criterion in distinguishing non-adiabatic from adiabatic perturbations will therefore simply be $\delta p_{ent} \ne 0$.
We therefore group together purely isocurvature perturbations (for which $\delta \ep = 0$ and hence $\delta p = \delta p_{ent}$) and arbitrary mixtures of isocurvature and adiabatic modes, using "non-adiabatic`` as an umbrella term to describe all overall pressure perturbations with non-zero $\delta p_{ent}$. It is worth emphasizing that this means we take "non-adiabatic'' to literally mean "not adiabatic``, and not only as a reference to purely isocurvature modes.

For notational convenience we also define an "effective`` speed of sound $\ce$ here, which relates pressure and energy density perturbations and reduces to $\ct$ when evaluated at constant entropy, i.e. in the adiabatic limit
\be
\ce = \frac{\delta p}{\delta \ep}.
\ee

\subsection{A perfect fluid}

The primary assumption going into our derivation will be treating the overall primordial matter as an effective perfect fluid (as it arises naturally for FRW universes).  Consequently our primordial matter stress-energy tensor will be of the following form
\be
{T^{\alpha}}_{\beta} = (\ep + p)u^{\alpha}u_{\beta} - p{\delta^{\alpha}}_{\beta}. \label{SE}
\ee
$u$ here refers to the potential 4-velocity.

Specific single fluid/field examples of matter with such a perfect fluid ${T^{\alpha}}_{\beta}$ include hydrodynamical fluids as well as scalar fields $\phi$ with Lagrangian $L_{\phi} = \sqrt{-g} p(X, \phi)$, where the kinetic term is $X = \frac{1}{2}g^{\alpha\beta}{\phi}_{,\alpha}{\phi}_{,\beta}$.  As such this incorporates canonical scalar fields with potential $V(\phi)$ as well as those with non-canonical kinetic terms (e.g. as encountered in k-inflation~\cite{garriga}). For a canonical scalar field $\phi$ we obtain \eqref{SE} with (see~\cite{mukh})
\be
p =  X - V(\phi) \hspace{1cm}    \ep = X + V(\phi).
\ee

More generally, however, we can view such an overall effective ${T^{\alpha}}_{\beta}$ as representing the sum of stress-energy tensors corresponding to (possibly multiple) components and their interactions. It is these interactions that can in principle give rise to non-adiabaticity in an overall perfect fluid.

To illustrate the mechanism by which this can happen, consider a perfect fluid with the following two-component decomposition
\be
{T^{\alpha}}_{\beta} = (\ep + p)u^{\alpha}u_{\beta} - p{\delta^{\alpha}}_{\beta} = {{T_1}^{\alpha}}_{\beta} + {{T_2}^{\alpha}}_{\beta}.
\ee
~\cite{Zimdahl:2006ng} then show that we can derive the following relationship independently of the nature of interactions between the two components
\bea
P - \frac{\dot{p}}{\dot{\ep}}D &=& \frac{\dot\ep_{1}}{\dot\ep}
\left(P_1 - \frac{\dot{p}_1}{\dot{\ep}_1}D_1 \right) +
\frac{\dot\ep_{2}}{\dot\ep}
\left(P_2 - \frac{\dot{p}_2}{\dot{\ep}_2}D_2 \right)\nonumber\\
&+& \frac{\dot\ep_{1}\dot\ep_{2}}{\left(\dot\ep \right)^2}
\left[\frac{\dot{p}_2}{\dot{\ep}_2} -
\frac{\dot{p}_1}{\dot{\ep}_1} \right] (D_2 - D_1)\ .
\label{Pnad}
\eea
Here $P$ and $D$ are measures of the overall fractional pressure and energy density perturbations, where subindices $i$ denote corresponding quantities for individual components $i$. 
\be
P_{i} = - 3H \frac{\delta{p}_i}{\dot\ep_i} \label{PA}
\ee
\be
D_{i} = - 3H \frac{\delta{\ep}_i}{\dot\ep_i} \ .\label{DA}
\ee

From our discussion above it is clear that $\frac{P}{D} = \ce$ and hence that $\frac{P}{D} = \ct = \frac{\dot{p}}{\dot{\ep}}$ in the adiabatic case. The first two terms in \eqref{Pnad} therefore vanish  if the individual components behave adiabatically. However, the third term is a measure of the interaction between the two components and is non-zero in general. From this we can immediately conclude that for the overall fluid $\frac{P}{D} = \ce \ne \frac{\dot{p}}{\dot{\ep}} = \ct$. In other words we can obtain non-adiabatic fluctuations in the overall perfect fluid in question. A closer look at the $\left[ \right]$ bracket term in \eqref{Pnad} also shows that departure from adiabaticity is directly linked to having distinct speeds of sound for individual components.

As an explicit example we here mention the familiar case of the Chaplygin gas, whose popularity stems from potentially providing an effective single fluid description of dark energy and dark matter. A Chaplygin gas has equation of state $p = \frac{A}{\ep}$. In its simplest forms $A$ is a constant, however here we will allow fluctuations in $A$ as well, following  ~\cite{Zimdahl:2006ng} once again. One then finds that the perfect fluid Chaplygin gas can be viewed as composed of another Chaplygin gas interacting with a pressureless fluid. Overall we obtain
\be
P = \ce D \hspace{1cm} \ce = \frac{\dot{p}}{\dot{\ep}} (1 - \mu)
\ee
\be
\mu = \frac{\ep}{(\ep + p) D} \frac{\delta A}{A}.
\ee 
For a non-zero $\mu$ we therefore again find $\ce \ne \ct$ and have produced a perfect fluid model with non-adiabatic perturbations.

\section{Perturbation Equations} \label{lin}

Here we derive the the equations of motion associated with relativistic primordial perturbations from the linearized Einstein equations. Establishing our notation we write the Friedmann equations as
\be
\Ha^2 = \frac{8 \pi G}{3} a^2 \ep   \label{F1}
\ee
\be
\Ha^2 - \Ha\pr = 4 \pi G a^2 (\ep + p) \equiv \Ga,  \label{F2}
\ee
where  $\ep$ and $p$ represent the overall background energy density and pressure respectively. We also have the conformal Hubble factor $\Ha \equiv a' / a$, $a(\eta)$ is the scale factor of the FRW metric and $\Ga$ has been defined in order to keep notation concise in what follows. Here $'$ denotes differentiation w.r.t conformal time $\eta$, which is defined as
\be
\int d \eta = \int \frac{dt}{a}.
\ee

The linearized perturbed Einstein equations for a perfect fluid ${T^{\alpha}}_{\beta}$ \eqref{SE} are 
\be
\Delta \Phi - 3\Ha(\Phi\pr + \Ha\Phi) = 4\pi G a^2 \db{\ep},
\label{scalar1}
\ee
\be
{{(a\Phi)}_{,i}}\pr = 4\pi G a^2 (\ep + p) \dbu = \Ga {\dbu}_i,
\label{scalar2}  
\ee
\be
\Phi^{''} + 3\Ha \Phi\pr + (2\Ha\pr + \Ha^{2})\Phi = 4\pi G a^2 \db{p}.
\label{scalar3}
\ee
Here $\Delta$ is the Laplacian, $\Phi$ is the Newtonian potential and the stress-energy fluctuations are evaluated in the longitudinal gauge (denoted by an overbar); we refer the reader to \cite{mukh, garriga} for notation and further explanations. We use equation (\ref{scalar2}) to solve for $\Phi^{'}$ and $\Phi^{''}$:
\be
\Phi\pr = \frac{\Ga}{a} \dbu - \Ha \Phi \label{phi'}
\ee
\be
\Phi^{''} = \frac{\Ga}{a}\dbu\pr + \frac{\Ga}{a}(\frac{p\pr}{\ep + p} - 3 \Ha) \dbu + \Ga \Phi  \label{phi''}
\ee
Combining equations (\ref{scalar1}) and (\ref{scalar3}) we get
\begin{equation}
\Phi^{''} +3(1 + \ce) \Ha \Phi\pr + (2\Ha\pr + (1 + 3\ce)\Ha^2 - \ce \Delta)\Phi = 0 \label{PhiEOM}
\end{equation}
Using (\ref{phi'}) and (\ref{phi''}) to substitute in for $\Phi^{'}$ and $\Phi^{''}$ we also derive the following equation for $\dbu$
\begin{equation}
\dbu\pr + \dbu (\frac{p\pr}{(\ep + p)} + 3 \ce \mathcal{H}) = (1 + \frac{\ce \Delta}{\Ga})(a\Phi)  \label{ueq} 
\end{equation}
Equations \eqref{PhiEOM} and \eqref{ueq} are effectively the equations of motion for the system under consideration, governing the Newtonian potential $\Phi$ and the matter 4-velocity $u$ respectively. They are also the precursors to the so-called $u$- and $v$-equations we encounter in the literature. 

Noticeably the $\dbu$ term in \eqref{ueq} vanishes for adiabatic perturbations, since
\be
\frac{p\pr}{\ep + p} + 3 \ct \Ha = 0.  \label{adf}
\ee
In the adiabatic case the equations of motion therefore simplify considerably. However, since we will eventually want to drop the requirement of adiabaticity, we proceed differently here. We combine the $\dbu\pr$ and $\dbu$ terms by introducing the new variable $\zetat \equiv \dbu f$, requiring 
\be
\frac{f\pr}{f} = \frac{p\pr}{\ep + p} + 3 \ce \Ha.   \label{fdef}
\ee
Importantly $f$ is therefore only defined up to its fractional variation here. Comparison with \eqref{adf} shows that $f$ provides a generic measure of the departure from adiabaticity. If we furthermore define $\xit \equiv a\Phi$, we have compactified our equations of motion \eqref{PhiEOM} and \eqref{ueq} to the following form
\be
\xit\pr = \frac{\Ga}{f}\zetat   \label{eom1}
\ee
\be
\zetat\pr = f(1 + \frac{\ce \Delta}{\Ga})\xit.   \label{eom2}
\ee
Using the methods outlined in Appendix A we can now also write down an action for this system
\be
S = \int({{\tilde{v}}}^{'2} + (\Ga + \ce \Delta)  {\tilde v}^2 + \frac{{\tilde z}^{''}}{{\tilde z}} {\tilde v}^2)d\eta d^3 x, \label{firstA}
\ee
where $\tilde v \equiv \tilde z \zetat$ and we have defined
\be
{\tilde z}^2 \equiv \frac{\Ga \hat{O}}{f (\Ga + \ce \Delta)}, \label{znaive}
\ee
where $\hat{O} (\Delta)$ is some time-independent operator (see appendix A for details). 
Whilst \eqref{firstA} is a perfectly valid action, however, it is representationally rather opaque. For instance, multiple terms here depend on $\Delta$ and hence (after Fourier-transformation) on the wave mode $k$. This obscures physically significant features such as the behavior of long- and short-wavelength modes, so that we will find it helpful to recast \eqref{firstA} into a form where such features as well as the relative contributions of entropic and adiabatic modes become more apparent. The next two sections are dedicated to this task, first in the adiabatic limit and then in the non-adiabatic case.

\section{The adiabatic case} \label{adia}

\subsection{General solution}

Here we are considering the limit where $\delta p_{ent} = 0$ and hence $f' = 0$. Using the techniques outlined in Appendix A we transform the equations of motion \eqref{eom1} and \eqref{eom2} by switching to new variables 
\be
\xx = \frac{a}{\Ha}\xit   \hspace{1cm} \xz = \frac{\Ha}{a}\zetat + \frac{1}{a}\xit.
\ee
In terms of these new variables we obtain equations of motion  
\be
\xx\pr = \frac{\Ga a^2}{\Ha^2}\xz  \hspace{1cm}  \xz\pr = \frac{\Ha^2}{\Ga a^2}\ct\Delta \xx.
\ee
The associated action is
\be
S = \int({v^{'}}^{2} + \ct \Delta  v^2 + \frac{z^{''}}{z}v^2)d\eta d^3 x,
\ee
where $v$ is the familiar canonical quantization variable $v = z \xz$ and
\be
z^2 \propto \frac{\Ga a^2 \hat{O}}{\Ha^2 \ct \Delta} \label{zad}.
\ee
Identifying the time-independent operator $\hat{O}$ with the Laplacian, $\hat{O} = \Delta$, we have reproduced the results obtained in the literature for adiabatic fluctuations~\cite{mukh}. $z^2$ is independent of $k$ and the representational advantages discussed above are restored. Furthermore $\zeta$ is the well-known curvature perturbation, which freezes out in the long-wavelength limit, where we can ignore the $\ct$ term. This feature is easily visible, once we've varied $S$ w.r.t $v$ and Fourier-transformed the resulting so-called $v$-equation, obtaining
\be
v^{''} + (\ct k^2 - \frac{z^{''}}{z}) v = 0. 
\ee

\subsection{A single scalar field}

Perhaps the most popular primordial matter candidate in the literature is a single scalar field. We will here summarize the argument given in~\cite{Christopherson:2008ry}, showing that a single $p(X,\phi)$ scalar field only produces adiabatic perturbations on large scales, as identically $\delta p_{ent} = 0$ for any such field then. For we can write the entropic pressure $\delta p_{ent}$ as
\bea
\label{Pent_gen}
\delta p_{ent} = \Big[&p_{,\phi}\left(1+\cs\right)-2\cs X p_{,X\phi}
\Big]\delta \phi + \nonumber \\ \Big[&p_{,X}\left(1-\cs\right)-2\cs X p_{,XX}
\Big]\delta X\,,
\eea
In addition we can derive the following constraint equation on large (super-horizon) scales
\bea
&\left(p_{,X}+\dot{\phi}^2 p_{,XX}\right)\delta X
+  \nonumber \\ &\left(\dot{\phi}^2 p_{,X\phi}-p_{,\phi}
+3Hp_{,X}\dot{\phi}\right)\delta \phi=0 \,,
\eea
Substituting this into the background field equations leads to $\delta X = \ddot{\phi} \delta \phi$. Finally, substituting this and the following expression for $\ct$
\be
\label{eq:csgeneral}
\cs=\frac{p_{,X}\ddot{\phi}+p_{,\phi}}
{p_{,X}\ddot{\phi}-p_{,\phi}+p_{,XX}\dot{\phi}^2\ddot{\phi}+
p_{,X\phi}\dot{\phi}^2} \,.
\ee
into \eqref{Pent_gen}, we find that
\be
\delta p_{ent} = 0.
\ee
A single $p(x,\phi)$ scalar field therefore only produces adiabatic perturbations on large scales, i.e. $\ce = \ct$.

\section{The non-adiabatic case} \label{nadia}

\subsection{General solution}

What happens if we allow $\delta p_{ent}$ to depart from $0$? Let us firstly recast the equations of motion into a form, where we separate out dependence on adiabatic and isocurvature modes. In terms of the adiabatic variables $\xz$ and $\xx$ we now obtain
\be
\xx\pr = \frac{\Ga a^2}{\Ha^2}\xz    \label{xxpr}
\ee
\be
\xz\pr = \frac{\Ha^2}{\Ga a^2}\ce\Delta \xx + \frac{f\pr}{f}  (\frac{\Ha}{a^2}\xx -  \xz).  \label{xzpr}
\ee
This explicitly reproduces the adiabatic equations in the $f' = 0$ limit and we can identify the last two terms in \eqref{xzpr} as ``non-adiabatic'' correction terms. Their dependence on $\ffpt$ shows that they vanish identically in the adiabatic limit. We have therefore separated out adiabatic and isocurvature modes at the level of the system's equations of motion.

In order to extract information about the behavior of asymptotic wave modes, it is, however, helpful to combine those separate contributions into a more compact form. Following this agenda, equations \eqref{xxpr} and \eqref{xzpr} can be further compactified by iteratively applying substitution schemes in analogy to the derivation of equations \eqref{eom1} and \eqref{eom2}.
After a series of such steps we have mapped variables $\xz$ and $\xx$ to 
\be
\theta \equiv \mu f \xz -  \nu \xx  \hspace{1cm} \rho \equiv \frac{\xx}{\mu}.     \label{thetadef}
\ee
Here the new substitution functions $\nu$ and $\mu$ are defined via coupled differential equations 
\footnote{Subject to the additional constraint (on the integration constants for $\mu$ and $\nu$) that $f\mu^2 \to 1$ as $f\pr \to 0$, i.e. ensuring the correct adiabatic limit.}
\be
\nu\pr = \frac{\Ha f\pr}{a^2} \mu   \hspace{1cm}  \mu\pr = \frac{\Ga a^2}{\Ha^2 f} \nu.    \label{mu}
\ee
With this new set of functions at our disposal we can finally write down the resulting equations of motion for the system
\be
\rho\pr = \frac{\Ga a^2}{\Ha^2 f}\frac{1}{\mu^2}\theta
\ee
\be
\theta\pr = \frac{\Ha^2 f}{\Ga a^2} \mu^2 \ce \Delta \rho,
\ee

Using techniques from Appendix A we can turn these equations of motion into an action 
\be
S = \int({v^{'}}^{2} + \ce \Delta v^2 + \frac{z^{''}}{z}v^2)d\eta d^3 x,
\ee
where the associated curvature variables $v$ and $z$ are given by
\be
z^2 \propto \frac{\Ga a^2}{\Ha^2 f \ce \mu^2}  \hspace{1cm}  v = z\theta. \label{newz}
\ee
Here we have once again identified $\hat{O}$ with the Laplacian $\Delta$.

With these variables the equation of motion for $v$ is formally identical to that derived for the adiabatic case
\be
v^{''} - \ce \Delta v - \frac{z^{''}}{z}v = 0. \label{vDelta}
\ee
All the effects of non-adiabaticity have been absorbed into $\ce$ and new expressions for $z^2$ and $v$ \eqref{newz} (and consequently a new time-dependent mass term $\zfppt$).

\subsection{Asymptotic limits}

Let us first consider the short-wavelength limit. Here the asymptotic solution to the fluctuation equations remains unaltered. To see this we Fourier-transform the $v$ equation, obtaining
\be
v^{''} + (\ce k^2 - \frac{z^{''}}{z}) v = 0. \label{vfour}
\ee
The two $v$-terms directly yield the short- and long-wavelength plane wave perturbations respectively. In the limit where the pressure term dominates (and hence $\ct k^2 >> |\zfppt|$) we can use the WKB approximation to give the short-wavelength solution 
\be
v \approx \frac{1}{\sqrt{c_{es} k}} exp(\pm i k \int c_{es} d \eta),   \label{vshort}
\ee
The functional form of the asymptotic short-wavelength limit is therefore independent of $\zfppt$ and hence remains unaffected by the introduction of non-adiabatic modes.

The long-wavelength solution is obtained by considering the opposite limit when the ``time-variable mass'' term dominates the pressure term (i.e. when $\ct k^2 << |\zfppt|$). The introduction of entropy modes therefore plays a significant role.
\be
v = C_1 (k) z + C_2 (k) z \int \frac{d\eta}{z^2} + O((k \eta^2)), \label{vlong}
\ee
where $C_1(k)$ and $C_2(k)$ are time-independent functions of the wavenumber $k$. Typically the first term in this expression represents the growing and therefore dominant mode allowing us to ignore the second term. We will discuss the effects of a dominant $C_2$ term in \ref{pertrev}. For a dominant growing $C_1$ mode, however, one immediately obtains a conserved quantity on super-horizon scales, i.e. in the long-wavelength limit. Following our notation above this is given by $\theta$, for $v = z\theta$ (see \eqref{newz}), thus resulting in a simple conservation equation $\theta\pr = 0$. This reduces to the familiar $\xz\pr = 0$ in the adiabatic limit. The ``frozen-in'' quantity outside the horizon, still formally given by $v/z$, therefore generalizes in the way presented. 

Relating the new expression for $\theta$ to the adiabatically conserved $\xz$ one finds
\be
\xz\pr = \frac{(\nu \xx)\pr}{f \mu} - (\frac{\mu\pr}{\mu} + \frac{f\pr}{f}) \xz.
\ee
The right-hand side of this equation can now be interpreted as an entropy producing source-term causing a deviation from the previously conserved $\xz$. This has important physical consequences. For example non-linearities could now continue to grow outside the horizon, potentially generating large levels of non-Gaussianity~\cite{piazza}.

Finally we can also obtain a general expression for horizon crossing modes, when considering the point where the short- and long- wavelength solutions ``meet''. Modes crossing from inside the Hubble radius to the gravity dominated region outside are then subject to the requirement that
\be
(k^2)\pr = {\left(\frac{z''}{\ce z}\right)}'>0.  \label{hcross}
\ee
If this inequality is not satisfied, then the modes considered are ``moving'' in the opposite direction, i.e. crossing the Hubble radius from the outside in.

\section{Cosmological constraints} \label{horiz}

The formalism developed in the previous sections is completely general, as long as our minimal set of assumptions is met. In particular we have not constrained the equation of state of the underlying fluid or the associated behavior of the scale factor. Inflating or non-inflating ($\ddot{a} >$ or $< 0$) as well as expanding and contracting ($\dot{a} >$ or $< 0$) solutions are still all on the table. Here we will compile a simple checklist to classify solutions arising from any particular model we might consider. 

\begin{itemize} 
\item Is the model capable of resolving the horizon problem? 

\item Is an ``expanding'' solution with $\dot{a} > 0$ possible? 

\item Can a scale-invariant power spectrum of perturbations be produced? The observed near scale-invariance of fluctuations~\cite{Komatsu:2008hk} therefore completes our check-list with an observational criterion.

\end{itemize}

Let us briefly expand on the items in this list. The horizon problem is posed by the fact that the scales we now observe are initially causally disconnected according to the unreformed Big Bang model yet are also stunningly uniform. In other words, the remarkably homogeneous and isotropic energy density distribution we observe inside the present horizon scale $c t_0 \sim 10^{28} cm$ appears to require a strongly fine-tuned initial matter distribution in the absence of a physical mechanism generating such uniformity. Schematically a "resolution of the horizon problem'' involves a temporary reversal of the kinematics realized in the unreformed Big Bang model. Specifically this requires the realization of a phase in the early universe when modes cross from the region inside the Hubble radius (where they are ruled by causal micro-physics) to the region outside (where they become dominated by gravity). With respect to the formalism this means that \eqref{hcross} needs to be satisfied by $k$-modes during this phase. 

In this context inflation is often invoked~\cite{infl}, where initial vacuum quantum fluctuations inevitably present while the modes are inside the Hubble radius are stretched to super-horizon scales during a phase with accelerating scale factor $\ddot{a}(t) > 0$. Typically this is achieved via the introduction of one or multiple scalar fields. Examples include slow-roll inflation, where restrictions are placed on the form of the scalar field's potential $V(\phi)$ (see e.g. \cite{Starobinsky:1980te, Stewart:2001cd}) and a variety of models with non-canonical kinetic terms (see e.g.~\cite{garriga, Silverstein:2003hf} and references therein for k-inflation and DBI-inflation respectively). Many alternatives exist, however: Cyclic or ekpyrotic scenarios~\cite{ekp, Khoury:2009my} propose a contracting pre-Big Bang phase ''causally connecting`` fluctuation modes. String gas cosmology~\cite{hag} invokes thermal fluctuations that exit the Hubble radius during an early quasi-static Hagedorn phase. Varying speed of light (or sound) frameworks~\cite{csdot, ArmendarizPicon:2003ht, Piao:2008ip} spread perturbations via a much larger speed of sound in the early universe, thus allowing all observed modes to originate from causally connected regions. Combinations of all of the above (see for example~\cite{piazza}) have also been proposed. More recently yet another potential resolution has been pointed out within Ho\v{r}ava-Lifshitz gravity~\cite{Horava:2009uw, Mukohyama:2009gg, Calcagni:2009ar, Gao:2009ht, Wang:2009yz}. Here the effective speed of light (and hence also the speed of sound) diverges in the ultraviolet regime, thus resolving the horizon problem in a way conceptually analogous to varying speed of light models.

The majority of these proposed ``resolutions`` to the horizon problem are realized within an expanding phase of the universe ($\dot{a} > 0$). However, there are exceptions; e.g. ekpyrotic scenarios or some of the models considered by us in~\cite{PriFluc}. Here the evolution is reversed as a contracting universe is getting denser and hotter in time rather than diluting and cooling. Features such as scale-invariant spectra are set up within a contracting phase in these models and the horizon problem is solved by fiat as causal connectivity is invariably produced during the contraction. In section \ref{pertrev} we point out some of the issues one must address when employing a contracting phase to generate fluctuations. 

One should note that there also remains a gauge problem in dealing with perturbations here. This is due to the fact that varying $w$ models give rise to different scales for metric and matter variables. We will not replicate the discussion of this topic here, but refer the interested reader to~\cite{PriFluc}.

\section{A model example: The Chaplygin gas revisited} \label{model}

We will now provide an example of how the formalism developed can be used to extract features for non-adiabatic perturbations around some specified background. The particular example we choose is an extension of the model discussed in~\cite{PriFluc}. Here the equation of state $w$ and the effective speed of sound $\ce$ exhibit a power-law dependence on the energy density $\ep$. In principle $w$ and $\ce$ of course do not have to be functions of $\ep$ only, but for illustrative purposes we will restrict ourselves to this simple model here. Other models exhibiting such a dependence include e.g. the Chaplygin gas and its modifications~\cite{chap}, but also intermediate inflationary models~\cite{barr,and}. In fact the latter leads to a second solution for scale-invariance in the inflationary setting (the other solution being slow-roll inflation~\cite{staro, Stewart:2001cd}).

\subsection{The background solution} \label{backg}

We consider models with equation of state $w \equiv {p}/{\ep}$ and effective speed of sound $\ce$ given by
\be
w + w_0 \propto \ep^{2\beta} \hspace{2cm} \ce \propto \ep^{2\alpha}   \label{backeq}
\ee
In the adiabatic limit $\ce = \ct$, which corresponds to $\alpha = \beta$ here. Specifying the equation of state in the adiabatic case will therefore automatically fix $\ct$ as well, so that we can then completely describe the solution via one parameter. In contrast, in the non-adiabatic case $\ce$ in principle becomes a free variable, so that we need at least two parameters to specify our model.

We will discuss solutions with positive and non-zero $\beta$ and positive $\alpha$ here. For other choices of $\alpha, \beta$ we refer to appendix B. At high density these models all display the same (power-law) behavior. A high $w$ phase then exits into a constant $w$, low density phase which need not be inflation. The corresponding low energy equation of state is given by $w_0$. We can also consider contracting models with these equations of state: the obvious generalization of cyclic models, where a constant high $w$ is invoked.

Solving for $\ep$ we get
\be
\ep \propto t^{\frac{-2}{1 + 4 \beta}}, \label{dens}
\ee
where we have ignored the low-energy contribution $w_0$. The energy density therefore changes like a power-law in $t$ and diverges as $t\rightarrow 0$. As such we have 
\be
w \propto t^{\frac{-4\beta}{1 + 4 \beta}} \hspace{2cm} \ce \propto t^{\frac{-4\alpha}{1 + 4 \beta}} 
\ee
Furthermore we get the following expression for the scale factor $a(t)$
\be
a(t) = a_0 e^{K_a t^{\frac{4\beta}{1+4\beta}}} \approx a_0.
\ee
where the approximate equality holds in the regime we shall be interested in, namely for early times $t<<1$.  The universe therefore appears to be loitering as a function of time ($a\approx a_0$) and $a$ does not vanish at $t=0$. This means we can effectively use conformal and proper time interchangeably, as they only differ by a constant factor: $\eta = t/a_0$ when $t << 1$. Nevertheless the fact that the universe's energy density diverges as $t \to 0$ shows that we do still have a Big Bang singularity. Completing our background solution we find that the conformal Hubble factor behaves as
\be
\Ha \propto a t^{\frac{-1}{1 + 4 \beta}}  \approx a_0 t^{\frac{-1}{1 + 4 \beta}}. \label{Hback}
\ee

\subsection{Perturbations and Power Spectra} \label{pert}

Having specified an equation of state $w$ and a particular profile for $\ce$ we are now in a position to explicitly solve the fluctuation equation \eqref{vfour}. In order to work out the associated power spectra of fluctuation modes we in principle need to know the full Bessel function solution for $v$. However, we can in fact extract all the necessary information from the asymptotic solutions to the $v$ equation \eqref{vshort} and \eqref{vlong}. When the growing ``$C_1$''-mode dominates, we can simplify these expressions for long- and short-wavelength modes respectively to the following form
\be
v \propto C_1(k) z  \hspace{1cm} v \propto \frac{1}{\sqrt{c_{es} k}},
\ee
where we have ignored a phase for the short wavelength limit. We can now determine $C_1 (k)$ by requiring these two asymptotic solutions to match when the $v$-term vanishes, i.e. at the point of horizon-crossing. Consequently we ``glue'' solutions at the point where
\be
\ce k^2 = \frac{z\pp}{z}.
\ee
The scalar power spectrum for $v$, $\Psp$, is given by
\be
\Psp \propto k^3 |\theta|^2,  \label{powerspectrum}
\ee
which we can express at the point of gluing as
\be
\Psp \propto k^3 |C_1 (k)|^2 \propto \frac{k^2}{z^2 c_{es}} = \frac{z\pp}{z^3 {c_{es}}^3}. \label{amp}
\ee
For a scale-invariant solution $\Psp$ has to be dimensionless. In other words $|C_1 (k)|^2 \propto k^{-3}$. Furthermore the dimensionless amplitude of the power spectrum is experimentally well constrained. We should also point out that \eqref{amp} is valid for modes exiting the horizon at a primordial stage regardless of the form of the long-wavelength solution for $v$. However, for it to still be a valid expression for modes re-entering the horizon at a later stage we require a dominant $C_1$ mode (see section \ref{pertrev} for details).

Substituting from the background solution into the definition of $z^2$ \eqref{newz} and differentiating, we also find that
\be
\frac{z\pr}{z} \propto \frac{4(\alpha - \beta)}{1 + 4\beta} t^{-1} - \frac{f\pr}{f} - 2\frac{\mu\pr}{\mu} + 2\frac{a\pr}{a}. \label{zpr}
\ee
This is interesting, as it contains information about which terms will dominate the evolution. Note that from \eqref{Hback} we can see that the first term will generically suppress the $a\pr/a = \Ha$ contribution, unless $\alpha = \beta$, i.e. the evolution is adiabatic. As such we can typically ignore the contribution of the $\Ha$ term in non-adiabatic cases. This is intriguing, given that it is the $\Ha$ term that provides the dominant contribution in familiar (adiabatic) inflationary models where $z^2 \propto a^2$. With an approximately constant $a_0$, however, we still obtain $\zfppt \propto {\eta}^{-2}$ when only the $t^{-1}$ term is significant, which {\it is} reminiscent of the analogous $\zfppt \propto {\eta}^{-2}$ expression we frequently find in inflationary models~\cite{piazza}.

\subsection{Adiabatic scale invariance} \label{pertad}

As can be seen from \eqref{backeq}, adiabaticity corresponds to $\alpha = \beta$ (with appropriately tuned proportionality constants). We here recover the familiar $z^2 \propto a^2$, which results in the following expression for $k$-modes in the early time limit
\be
(k^2)\pr \propto (\Ha^2)'<0.
\ee
No expanding resolution of the horizon problem is therefore possible here, as modes are crossing over from outside the Hubble radius to the inside region. This is the opposite behavior to the one necessary to realize the desired reversal of the unreformed Big Bang model kinematics. In order to address the horizon problem this model consequently has to be implemented within a contracting phase, as e.g. invoked in cyclic scenarios. For in such a phase the evolution is reversed and incoming modes are mapped into outgoing ones and vice versa.

Applying the gluing techniques discussed above we can now compute the scalar power spectrum for adiabatic perturbations at the point of horizon-crossing. We find
\be
\Psp \propto t^{\frac{2\beta-2}{1+4\beta}},
\ee
straightforwardly giving us a criterion for scale invariance by requiring the spectrum to be $k$- and hence, at horizon-crossing, time-independent
\be
\beta = \alpha = 1.
\ee
This result is in agreement with the one obtained in~\cite{PriFluc}. We therefore recover a scale-invariant solution which furthermore is capable of solving the horizon problem in a collapsing phase. We should note that by simultaneously requiring adiabaticity and $\beta > 0$ we have only explored time-varying $\ct$ solutions here.

\subsection{Non-adiabatic scale invariance I}

Let us first consider non-adiabaticity with $\alpha > \beta$. Comparing with \eqref{zpr} we then find 
\be
\frac{z\pr}{z} \propto \frac{f\pr}{f} \propto t^{\frac{-4\alpha -1}{4\beta + 1}} \label{zprnad}
\ee
As such $f$ will exhibit an exponential dependence. We again emphasize that the usually important contributions from the scale factor are negligible here. We are therefore in a radically different regime from the standard inflationary $\zfpt \propto \afpt$ scenario. With this the fluctuations are fully specified and, after computing the time-variable mass term $\zfpp$ to first order, we find
\be
(k^2)\pr \propto (t^{\frac{-4\alpha - 2}{4\beta + 1}})'<0. \label{kred}
\ee
Analogously to our reasoning for the adiabatic case, we can thus conclude that no expanding resolution of the horizon problem is possible here as well. In fact this conclusion is not an artefact of choosing a power-law solution for $\ce$, but will obtain for a much wider class of $\ce$ profiles, e.g. if an exponential trial solution $\ce \propto e^{k t^{-c}}$ is chosen for some constants c and k. In the class of scenarios considered here it would therefore require a change in the equation of state to overcome the solution's ``affinity'' towards collapsing models. For, whilst an initially large and subsequently decaying $\ce$ has the tendency to reverse the horizon crossing behavior of k-modes, the time variable mass term $\zfppt$ counteracts this here.

Moving on to computing the scalar power spectrum we obtain
\be
\Psp \propto \frac{z\pp}{z^3 {c_{es}}^3} \propto \frac{t^{\frac{-2\alpha -2}{1 + 4 \beta}}}{z^2}.
\ee
However, the exponential dependence of the $f$-measure \eqref{zprnad} here means that $z^2$ diverges as $t \to 0$. And hence $1 / z^2$ vanishes as $t \rightarrow 0$, growing with time afterwards. From expression \eqref{amp} we see that the amplitude of the power spectrum $\Psp$ will consequently grow with time, also showing divergent behavior. Comparing this with \eqref{kred} we find that the result is an exponentially red-tilted power spectrum. Baring extreme fine-tuning, no (near)scale-invariant solution can therefore be produced here.

\subsection{Non-adiabatic scale invariance II}

We will now consider the non-adiabatic regime with $\beta > \alpha$. Prima facie one would not expect this regime to succeed in stretching primordial fluctuations to super-horizon scales in an expanding phase. For whilst we can still have an initially large speed of sound here, its effects are offset by a more rapidly varying equation of state $w$. Nevertheless it might present an interesting possibility for contracting scenarios. Let us first back up the claim that no expanding resolution of the horizon problem is possible in this context though.

Regarding the formalism we find that the ``non-adiabatic'' functions $f$ and $\mu$ are given by
\be
f \propto \Ga \hspace{1cm} \mu \propto \Ga^{-1}
\ee
For $\beta > \alpha \ge 0$ we consequently get horizon-crossing modes with
\be
(k^2)\pr \propto (t^{-2+\frac{4\alpha}{1+4\beta}})'<0. \label{kblue}
\ee
This does indeed show that no expanding solution can solve the horizon problem in the proposed fashion here. One should point out that the reasoning for the case where $\alpha = \beta$, but adiabaticity fails due to non-matching proportionality constants, is entirely analogous to the considerations presented in this section and the same equations are reproduced.

Is the $\beta > \alpha$ solution a candidate for a contracting solution to the horizon problem? Turning our attention to the scalar power spectrum we find that $\Psp$ at horizon-crossing is given by
\be
\Psp \propto t^{\frac{2\alpha}{1+4\beta}},  \label{powerbeta}
\ee
For a scale-invariant solution we require \eqref{powerbeta} to not be $t$-dependent, so that the condition for scale invariance is simply
\be
\alpha = 0.
\ee
This corresponds to having a constant effective speed of sound $\ce$. We have therefore found a non-adiabatic, contracting and scale-invariant solution with constant $\ce$ but arbitrary $w \propto \ep^{2\beta}$ here. Interestingly it is straightforward to fabricate blue-tilted spectra in the $\beta > \alpha \ge 0$ regime. This can be achieved by choosing positive, non-zero $\alpha$ as can be seen from \eqref{powerbeta}. A red-tilted spectrum is not possible here, however.

In summary we have found a family of equations of state $w$ that satisfy our background solution with $\beta > \alpha$, which can resolve the horizon problem and produce flat power spectra in a contracting phase.

\section{Growing vs. Decaying modes} \label{pertrev}

We remind ourselves of the conservation equation for long-wavelength modes \eqref{vlong}
\be
v = C_1(k) z + C_2(k) z \int \frac{d\eta}{z^2} + O((k \eta^2)).    \label{vlong2}
\ee
This will only give rise to a conserved quantity $\theta$ if the $C_1$ term is the dominant/growing mode.
If the $C_2$ mode is dominant, on the other hand, $\theta$ is not conserved. A power spectrum scale-invariant in $\theta$ upon horizon exit consequently does not guarantee that modes re-entering the horizon at a later stage will be scale-invariant in $\theta$ anymore. Setting up a primordial scale-invariant power spectrum then no longer automatically justifies observing a scale invariant spectrum today. Meeting the observational constraints therefore requires a more elaborate setup than the simple gluing considerations we presented for a growing $C_2$ mode. As an example of such a scenario with dominant $C_2$ term we here refer to a single scalar field in a contracting phase (\cite{Khoury:2001zk, Tolley:2007nq}). 
Such setups frequently suffer from the problem of obtaining generically blue $\xz$ fluctuations (for an exception see \cite{Khoury:2009my}).

The solutions worked out in the previous sections must also be implemented in a collapsing phase if they are to resolve the horizon problem. It is thus natural to ask, whether we are also faced with a dominant $C_2$ mode here. For in constructing such a collapsing phase one effectively considers the mirror image of the post Big-Bang picture we are familiar with. 
As temporal behavior is reversed upon mirroring the dynamics around a singularity at $t=0$, growing modes can be mapped into decaying ones and vice versa. When implementing a solution in a contracting phase, one therefore needs to be careful to not ignore previously suppressed solutions that can become dominant in a collapsing phase.

For the example model presented here the following picture emerges:
In the adiabatic limit the $C_2$ term is proportional to $\int \frac{d\eta}{a^2} \propto |\eta|$. It consequently vanishes as $t \to 0$ and only the $C_1$ term stays relevant.

When $\alpha > \beta > 0$ the $C_2$ contribution exponentially tends to zero as $t \to 0$. This can be seen from the exponential dependence of $z^2$ \eqref{zprnad}. It therefore becomes suppressed and hence irrelevant in this limit. Thus we maintain a conserved super-horizon quantity $\theta$ even in the case of a collapsing phase.

When $\beta > \alpha > 0$ we find that the $C_1$ and $C_2$ modes are proportional to
\be
z_1 \propto t^{\frac{2\alpha - 4\beta - 1}{1 + 4\beta}} \hspace{1cm} 
z_2\int \frac{d\eta}{{z_2}^2} \propto t^{\frac{-2\alpha + 8\beta +2}{1+4\beta}}.
\ee
From these expressions it again follows that the $C_1$ mode does indeed dominate over the $C_2$ mode and we therefore maintain a frozen-in $\theta$ on super-horizon scales. 

We have therefore justified that the ``gluing'' procedure in our derivation does give us the correct power spectra both at horizon exit and re-entry for all cases considered. In other words, we are protected from additional super-horizon fluctuations sourced by $C_2$ modes. We will not discuss the issue of matching modes across the singularity here, but have worked under the assumption that perturbations set up in a contracting phase can be successfully mapped into a post Big-Bang expanding phase (for a more detailed discussion in the ekpyrotic setting see e.g.~\cite{Durrer:2001qk,Khoury:2001zk,Lyth:2001pf}).

\section{Conclusions} \label{conc}

We considered general non-adiabatic cosmological perturbations, allowing for varying $w$ and $\ce$. In this setting we derived the second-order action and its curvature variables in the same generality as has been achieved for purely adiabatic modes before. This showed how non-adiabatic perturbations give rise to a new conserved super-horizon quantity $\theta$. Relating it to the adiabatically conserved $\xz$ draws attention to potential consequences such as enlarged non-Gaussianities.

Applying the new formalism to an extension of the specific background model discussed in ~\cite{PriFluc}, we then produced both an adiabatic and a non-adiabatic scale-invariant solution for structure formation. The specific model considered has to be implemented in a contracting phase in order to resolve the horizon problem. In this context we considered the issue of growing and decaying modes.  We show that, untypical for contracting models, the $C_1$ mode dominant in the corresponding expanding case also stays dominant here.

It will be an interesting task for the future to investigate whether models with other equations of state and speed of sound profiles can produce further expanding and (near) scale invariant scenarios for structure formation. We hope the formalism developed will be useful in this enterprise.

\bigskip
{\bf Acknowledgements} We are grateful to an anonymous referee for very helpful comments. JN would like to thank Larry Ford, Justin Khoury, Alexander Vilenkin and Daniel Wesley for insightful discussions. This work was supported by an STFC studentship.

\section{Appendix A: Constructing a schematic action} \label{scheme}

Consider a system with equations of motion:
\begin{align}
\xi\pr + A(\eta)\xi &= B(\eta)\zeta&
\zeta\pr + C(\eta)\zeta &= D(\eta)\xi,
\end{align}
where $\eta$ is a ``time'' associated with the system, $'$ denotes differentiation w.r.t. $\eta$ and ($\xi$,$\zeta$) are the variables of the system.
We want to find an action which, when varied w.r.t. $\xi$ and $\zeta$, reproduces these equations of motion. 

For this consider the following toy Lagrangian
\begin{equation}
L = \alpha(\eta)\xi\hat{O}\xi + \beta(\eta) \zeta\hat{O}\zeta + \gamma(\eta) \xi\hat{O}\zeta + \delta(\eta) \zeta\hat{O}\xi\pr + \sigma(\eta) \xi\hat{O}\zeta\pr, \label{GenLag}
\end{equation}
where $\hat{O} = \hat{O}(\Delta)$ is some time-independent operator.
The Euler-Lagrange equations then allow us to fix the $\alpha(\eta) .. \sigma (\eta)$ terms. Furthermore one can eliminate the $A(\eta)$ and $C(\eta)$ terms by employing substitution schemes in analogy with the way we defined the $f$-measure~\eqref{fdef}. Without loss of generality and with an eye on equations  \eqref{eom1} and \eqref{eom2} we therefore set $A(\eta)$ and $C(\eta)$ to zero here. In fact this procedure gives us a simple tool to combine e.g. terms linear in a variable $\xi$ with terms linear in $\xi\pr$ via a change of variables. After some algebra we then obtain the following Lagrangian 
\begin{equation}
L = {\frac{K_1}{2}D(\eta)\xi\hat{O}\xi - \frac{K_1}{2}B(\eta)\zeta\hat{O}\zeta - K_1\xi\hat{O}\zeta\pr},
\end{equation}
where $K_1$ is a constant of integration and we have eliminated an extra degree of freedom by choosing $\sigma = - K_1$.
Replacing $\xi$ via \eqref{eom2} we can therefore finally write down the action for the full system:
\be
S(\zeta, \zeta^{'}, \eta) = \int{z^{2}({\zeta^{'}}^{2} + B(\eta)D(\eta)\zeta^{2})d\eta d^{3} x},
\ee
where we have defined
\be
z^2 \equiv \frac{\hat{O}}{D(\eta)}.
\ee
 
We are now in a position to introduce the variable $v \equiv z\zeta$, for which the action (up to a total derivative) takes on the following form
\begin{equation}
S = \int({v^{'}}^{2} + B(\eta)D(\eta)v^2 + \frac{z^{''}}{z}v^2)d\eta d^3 x. \label{vact}
\end{equation}
Varying this action w.r.t. $v$ we obtain the equation of motion:
\begin{equation}
v^{''} - B(\eta)D(\eta)v - \frac{z^{''}}{z}v = 0.  \label{veom}
\end{equation}

\section{Appendix B : Regimes of the example model}

\subsection{A constant equation of state} \label{negcon}

Constant equations of state abound in the literature. They are invoked e.g. in the standard de Sitter solution or by cyclic models~\cite{Gratton:2003pe} where a constant, high $w$ is postulated. In our setup they correspond to $\beta = 0$. We therefore get
\be
1 + w = \gamma.
\ee
where $\gamma$ is a constant. If $\gamma = 0$, the familiar exponentially inflating de Sitter solution is reproduced
\be
a(t) \propto e^{K t},
\ee
If $\gamma > 0$, then the expression for the scale factor $a$ becomes
\be
a(t) = a_0 t^{\frac{2}{3 \gamma}}.
\ee
Consequently we are in an inflationary regime if $\gc(\gc-1) > 0$. In particular we have power law inflation for $\gamma < 2/3$. This model therefore either always shows accelerated expansion or it never does at all, depending on what the constant value of $w$ is tuned to.

The energy density is now given by
\be
\ep = \frac{4}{3 \gamma^2} t^{-2},
\ee
i.e. we obtain the well-known ``scaling solution'' for $\ep$. At $t = 0$ the scale factor goes to zero and the energy density diverges, i.e. we have a Big Bang singularity. 

If $\alpha$ is positive and non-zero, we obtain the following behavior of $k$-modes
\be
(k^2)' \propto (t^{-4\alpha -2 + \frac{4}{3 \gamma}})' < 0.
\ee
We therefore have both expanding and contracting solutions depending on how parameters are chosen. If $4\alpha + 2 > \frac{2}{3\gamma}$ we are in a contracting scenario. Conversely, if the inequality is not satisfied, an expanding model is the result. However, in both cases scale invariance fails to obtain by the same arguments as given in the $\alpha > \beta$ case. For we get an exponential contribution via the $f$ dependence of $z^2$ \eqref{zprnad}.

In the constant $\ce$ limit, on the other hand, (i.e. $\alpha = 0$) the condition for an expanding solution interestingly becomes $\gamma < 2/3$, so that we have expansion if and only if the model displays power law inflation. Such models are well known, so we will not discuss them further here.

It is worth pointing out that the considerations for $\beta = 0$ also apply to generic $\beta < 0$ models. Remember we are considering an equation of state of the form
\be
w + w_0 = \gamma \ep^{2 \beta}.
\ee
Combining this with the Friedmann equations we get
\be
\dot{\ep} + \sqrt{3}\ep^{\frac{3}{2}}(\gamma \ep^{2 \beta} - (w_0 -1)).
\ee
If $w_0 \not= 1$, the $\ep^{2\beta}$ term is suppressed for $\beta < 0$. Hence we again have a scaling solution for the energy density $\ep \propto t^{-2}$. Having gotten rid of a constant of integration by shifting the time coordinate, this takes on the following form
\be
\ep = \frac{4}{3} (w_0 -1)^{-2} t^{-2}.
\ee
For $\alpha = 0$ we are in the power law inflation regime, otherwise this results in an exponentially divergent spectrum. This case is therefore completely analogous to that of a constant equation of state considered above.

\subsection{Tuned $\beta < 0$ scenarios}

The cases where $\beta < 0$ and the equation of state is tuned to reproduce the standard perfect fluid equation of state, i.e. choosing $w_0 = 1$, are noticeably different. The corresponding adiabatic background solution is discussed at length in \cite{barr}. Essentially we reproduce the expressions presented in \ref{backg} unless $\beta = -\frac{1}{4}$. As a general feature such models are naturally drawn towards $w = -1$ either as $t \to 0$ or for $t >> 1$. The evolution is therefore either set up in an inflationary de Sitter phase or eventually exits into one.

If $0 > \beta > -\frac{1}{4}$, the constant $a$ assumption breaks down at first sight. However, recall that
\be
a = a_0 e^{K_a t^{\frac{4\beta}{1+4\beta}}}.
\ee
$K_a$ is now negative and $a$ is consequently strongly suppressed, i.e. $a << a_0$ at early times. With respect to the background solution this changes very little though. We recover $\ffpt \propto 3 \Ha \ct$ in analogy with the $\alpha > \beta$ case. Consequently no scale-invariant solution is possible just as in the $\alpha > \beta$ case, for $z^2$ is exponentially divergent.

If $\beta < -\frac{1}{4}$ the background behaves analogously to the $\beta > \alpha \ge 0$ case. $a$ is therefore loitering as a function of time. Consequently we can replicate the expression for horizon-crossing modes 
\be
(k^2)\pr \propto (t^{-2+\frac{4\alpha}{1+4\beta}})'<0.
\ee
Moving to negative $\beta$ has not affected the fact that the solution is contracting. On the contrary the contraction is more rapid now. The constant $\ce$ solution can still be implemented and we have identified the negative $\beta$ analogue of the non-adiabatic scale-invariant solution found in the positive $\beta$ domain. Importantly the consideration of growing and decaying modes for the $\beta > \alpha \ge 0$ case in \ref{pertrev} also carries over and we find a dominant $C_1$ mode again.

\end{document}